\documentclass[twocolumn,showpacs,preprintnumbers,amsmath,amssymb]{revtex4}

\usepackage{graphicx}
\usepackage{dcolumn}
\usepackage{bm}
\usepackage{color}

\begin{document}


\title{Diffusion between evolving interfaces}

\author{Janne Juntunen}
\email{janne.k.juntunen@jyu.fi}
\author{Juha Merikoski}%
 \email{juha.t.merikoski@jyu.fi}
\affiliation{%
Deparment of Physics, University of Jyv\"askyl\"a
}%


\date{24 September 2010}

\begin{abstract}
Diffusion in an evolving environment is studied by 
continuos-time Monte Carlo simulations.
Diffusion is modelled by continuos-time random walkers 
on a lattice, in a dynamic environment provided by 
bubbles between two one-dimensional interfaces driven symmetrically 
towards each other. For one-dimensional random walkers 
constrained by the interfaces, the bubble size distribution 
dominates diffusion. For two-dimensional random walkers, 
it is also controlled by the topography and dynamics 
of the interfaces. The results of the one-dimensional 
case are recovered in the limit where the interfaces 
are strongly driven. Even with simple hard-core repulsion 
between the interfaces and the particles, diffusion 
is found to depend strongly on the details of the 
dynamical rules of particles close to the interfaces.
{\it Article reference: Journal of Physics: Condensed Matter 22, 465402 (2010).}
\end{abstract}

\maketitle

\section{\label{sec:Introduction}Introduction}

Diffusion phenomena are ubiquitous in nature, familiar examples 
ranging from heat conduction to osmosis. Often diffusion 
occurs in a random or nonideal environment as it does, for example, 
in the presence of mobile or immobile (in the time scale of diffusion) 
lattice imperfections. Due to the complexity of real materials, 
transport has been considered within simplified theoretical frameworks, 
often utilizing the random-walk picture to describe some assumed 
underlying microscopy. Novel biophysical applications can be expected 
to emerge for transport restricted by soft (and fluctuating) interfaces 
forming narrow or even nanoscale channels \cite{Sparreboom2010}. 
In particular the crossover from bulk-dominated to boundary-dominated 
diffusion is of considerable theoretical and experimental interest.

The study of random walks in a random environment (RWRE) has
a long history and since the results~\cite{Solomon75,Kesten75} 
from the 70's, as reviewed in Ref.~\cite{Zeitouni06}, a vast 
amount of information has been accumulated. 
This randomness has been considered to manifest itself 
as non-homogeneous transition rates~\cite{Alexander81,Doussal99,Sadjadi08}.
Spatially, (frozen) transition rates can sometimes be described 
by random walkers like in the Sinai model~\cite{Sinai82}. 
In general, theoretical studies have mostly been limited to models, 
where the environment, including the possible geometric 
constraints, is either stationary or fast compared with the 
jump rate of the walkers. 
The mathematical problem of the random walk in an uncorrelated fluctuating 
environment has been considered in Ref.~\cite{Boldrighini2007}. 
The asymptotics of diffusion in continuum under a random forcing 
in the presence of damping were analyzed in Ref.~\cite{Mehlig09} and
diffusion in restricted geometries with homogeneous 
transition rates was considered in Ref.~\cite{Burada09}.
Two-species zero-range process \cite{Evans05,Spitzer70} 
with suitably chosen transition rates leads to dynamics which can be considered 
as a diffusing particle in an evolving environment ~\cite{Evans03}. 
However, in existing studies the focus has not been in (undriven) diffusion.  
In Ref.~\cite{Sane09} San$\acute{\textrm{e}}$ {\em et al.} considered 
a situation, where particles are immersed in a background fluid inside a narrow channel. 
From the point of view of a single particle in the dilute limit, 
this could be interpreted as diffusion in a dynamic environment. 
In that work the focus was on the transition from single file 
dynamics to Fickian diffusion. The existing studies on particle 
dynamics in the presence of interfaces are mainly for particles 
immersed in a driven liquid \cite{Kunert07}. A related problem, 
the influence of geometry fluctuations on lateral diffusion in 
biological systems was studied very recently in Ref.~\cite{Chevalier10}, 
where it is noted that geometry fluctuations at a finite scale 
can affect diffusion at all scales.

 
In this Article, we consider diffusion in a dynamic  
restricting environment, inside open evolving 'bubbles' 
between two interfaces. For this, we combine two most simple 
models, the solid-on-solid (SOS) model of interfaces 
and the continuous-time random walk on a lattice, 
models that are widely used and known to describe 
well interface fluctuations and particle diffusion.
To be more specific, we study diffusion on a lattice 
in the environment produced by the dynamics 
of the BCSOS2-model introduced in Ref.~\cite{Juntunen07}, 
containing two non-intersecting interfaces driven against 
each other. Thus, due to the interface dynamics, 
the actual transition rates of a diffusing particle 
become dependent on time and position, when the jumps of 
the particle are possible only inside the 'bubbles' 
between the interfaces. Diffusion in the 
hydrodynamic limit will then depend on the 
dynamics of the bubbles, e.g.~trough their growth and 
merging. We shall concentrate on cases, where the 
particles do not affect the motion of the interfaces 
so that the dynamics of the two interfaces, 
e.g.~the bubble-size distributions and their 
correlations, are in principle known \cite{Juntunen07} and thus, 
in addition to simulations, analytical arguments can be developed 
for various limits, which is the particular strength of the model. 
We use our model as a testing ground for various ideas 
describing different regimes of behavior of the diffusion 
coefficient. In addition, we study the possible consequences 
of various choices of microscopic dynamics on the interaction 
of the interfaces and the diffusing particle. 
To reduce the dimension of the parameter space, the 
models we combine are relatively simple. However, most 
of our results are expected 
not to be dependent on the details of the model, but are 
characteristic of systems, where the size distribution 
of the bubbles and their dynamics (at the bubble scale) inside 
a material or at an interface between two materials become 
the rate-limiting factor for diffusion (at the hydrodynamic scale).

This paper is organized as follows. In Sec.~\ref{sec:Models} 
we define the evolving environment provided by the two 
interfaces and the dynamic rules of the walker (particle) in detail. 
After that, in Sec.~\ref{sec:Numerics}, 
we describe the algorithms needed for efficient simulation of the combined 
dynamics and the sampling of the main quantities. 
An experimentally oriented reader could first 
skip Secs.~\ref{sec:Models}-\ref{sec:Numerics} and proceed to 
Sec.~\ref{sec:Parameters}, where we briefly describe the 
parameters and their physical interpretation. 
Our results for the combined interface and particle dynamics 
are presented in Secs.~\ref{sec:ResultsInt}-\ref{sec:ResultsDif}. 
A concluding discussion of our results is given in Sec.~\ref{sec:Conclusions}.

\begin{figure}[htbp!]
	\begin{center}
		\includegraphics[width=0.45\textwidth]{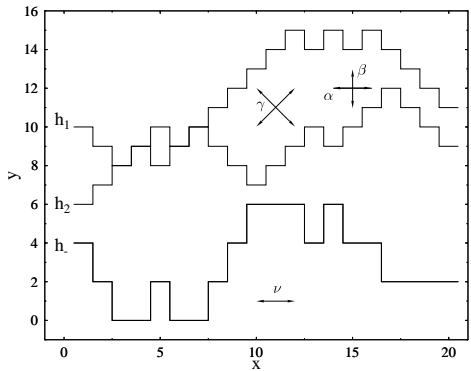}
	\end{center}
	\caption{A snapshot of the two interfaces $h_1$ and $h_2$ and the possible 
	jumps of a diffusing particle with their attempt rates 
	$\alpha$, $\beta$,  $\gamma$ and  $\nu$. 
	Also the corresponding difference $h_-=h_1-h_2$ is shown.
	The jumps of the particle are possible only within the bubbles (open spaces) 
	between the interface. The jump rate $\nu $ is for 
	the one-dimensional motion (see dynamics 4 in text) 
	depending only on $h_-$.  To simplify the discussion, when possible, in the text 
	the various rate parameters are collectively referred to simply as $\mu$ 
	in such a way that the total rate for the particle to attempt a jump is $\mu$.
		  }
\label{fig:figure1}
\end{figure}

\section{\label{sec:Models}Models}
\subsection{\label{sec:Interface}Model for the interface dynamics}

The evolving environment for diffusion is produced by the 
dynamics of the BCSOS2 model that we introduced and discussed 
in Ref.~\cite{Juntunen07}. 
Below we give only 
a brief description of the BCSOS2 model so that the dynamical 
rules for diffusion become well defined for the present study. 
The BCSOS2 model is 
constructed by letting two one-dimensional BCSOS 
interfaces \cite{Barabasi95} interact with each others.

The location or the 'height' of a single BCSOS interface is 
described by a function $h_i(x,t)$ such that, for every 
site $x=1,...,L$,
\begin{eqnarray}
  \label{eq:bcsos}
  h_i(x+1,t)-h_i(x,t)=\pm 1, 
\end{eqnarray}
where we without loss of generality assume 
that the possible values of $h_i$ are integers.
With this restriction on local configurations of the interfaces, 
only two kinds of processes, adsorption ($h_i$ locally increases) 
and desorption ($h_i$ locally decreases), are available. 
In our continuous-time model (see~Sec.~\ref{sec:Numerics}), 
the parameters $p_i$ and $q_i$ give the transition rates 
of adsorption and desorption events, respectively, 
for transitions allowed by the condition (\ref{eq:bcsos}). 
In what follows, time and various rates are 
measured in units where $p_i+q_i=1$.

In the BCSOS2 model there are two interfaces, $h_1(x,t)$ and
$h_2(x,t)$, such that $h_i(x,t)$ is an even (odd) integer for 
odd (even) values of $x$, see Fig.~\ref{fig:figure1}.
The coupling between the interfaces is produced by demanding 
that they cannot intersect: 
\begin{eqnarray}
  \label{eq:nocross}
  h_1(x,t)\ge h_2(x,t) \ \ \textrm{for all} \ \ x,t.
\end{eqnarray}
We also impose the periodic boundary 
conditions $h_i(x,t) \equiv h_i(x+L,t)$ for $i=1,2$. 
In the full BCSOS2 model, there are then four parameters 
$[(p_1,q_1),(p_2,q_2)]$ defining the transition rates for 
the interfaces $h_1$ and $h_2$, respectively. 
To further limit the parameter space, we shall restrict 
the discussion to the symmetric 
case $p_1=q_2$ and $q_1=p_2$ (see Ref.~\cite{Juntunen07}) 
so that the behavior of the BCSOS2 system is defined 
by one parameter, the driving parameter $f$ defined as
\begin{eqnarray}
  \label{eq:parameters}
  f \equiv p_2/q_2-1.
\end{eqnarray}
For large $f$ the interfaces are strongly driven against 
each other and for $f\to 0$ they become free.
We also define the sum and difference processes defined 
via $h_{\pm}=h_{1} \pm h_{2}$, where the sum process 
$h_+$ describes the wandering of the interfaces together 
and the difference process $h_-$ is positive inside 
the bubbles and zero elsewhere \cite{Juntunen07}. 
The non-crossing condition of Eq.~(\ref{eq:nocross}) 
is equivalent to $ h_{-}(x,t) \ge 0$.
The interfaces $h_{\pm}$ are then of the RSOS type 
\cite{Barabasi95}, obeying 
$  h_{\pm}(x+1,t)-h_{\pm}(x,t)= -2,0,+2 $.
An example of $h_1$ and $h_2$ and the 
corresponding $h_-$ configuration is shown 
in Fig.~\ref{fig:figure1}.

\subsection{\label{sec:Particles}Models for particle diffusion }

We consider a single point-sized particle diffusing between the 
interfaces $h_{1}(x,t)$ and $h_{2}(x,t)$ on a lattice $(x,y)$. 
The lattice point coordinates in the horizontal direction are 
the same $x=1,...,L$ as for the interface model above, again with 
periodic boundary conditions, c.f.~Fig.~\ref{fig:figure1}. 
In the 'vertical' direction, the lattice is infinite and the 
coordinates are integers $y=...,-2,1,0,1,2,...$ and thus coincide 
with the possible values of $h_1$ and $h_2$.

We shall denote the location of the particle by $(x_p,y_p)$. 
The particle does not affect the dynamics of the interfaces 
but, if needed, a moving interface can push the particle 
the distance of one or two lattice units in the vertical direction 
such that the location of the particle also after the change of 
the interface configuration 
satisfies the condition 
\begin{equation}
h_2(x_p,t) \le y_p \le h_1(x_p,t).
\end{equation}
These are moves of the particle forced by the interface motion.


For diffusive moves of the particle, the following two rules 
are imposed in all cases: 
First, for a jump $(x_p,y_p)\to(x_p',y_p')$ to be possible, 
the product of the interface height differences on the 
departure site and the arrival site is 
non-zero: $h_{-}(x_p,t) h_{-}(x_p',t)>0$ {\em i.e.} 
the channel for the jump between the interfaces must be open 
at both ends of the jump.
Second, an attempted jump arriving outside the region bounded by 
the interfaces is blocked.
In the actual dynamics, the direction of an attempted 
jump is chosen without any prior knowledge of the ability 
of the particle to perform the jump.

For diffusion on the square lattice there are a few natural choices 
for the possible particle jumps $(x_p,y_p)\to(x_p',y_p')$

{Dynamics $m=1$:} The most obvious case are the nearest-neighbor jumps such that the particle 
jumps in the  horizontal direction $(x_p,y_p) \to (x_p\pm 1,y_p)$ with the attempt rate $\alpha$ 
and in the vertical direction $(x_p,y_p) \to (x_p,y_p\pm 1)$ 
with the attempt rate $\beta$. We set $\alpha=\beta$ so that the total 
attempt rate of the particle is $\mu = 4\alpha$.

{Dynamics $m=2$}: The particle jumps diagonally, {\em i.e.} $(x_p,y_p) \to (x_p\pm 1,y_p\pm 1)$ 
independently, with the attempt rate $\gamma$. The total attempt rate is then $\mu=4\gamma$.
This process is expected to be efficient on tilted sections of the interfaces like 
the rightmost part of the snapshot configuration in Fig.~\ref{fig:figure1}.

{Dynamics $m=3$}:  This is a combination of the jumps available in dynamics 1 and 2. 
In this work we chose $\alpha = \beta = \gamma $ so that $\mu=8 \alpha$.

In addition to the three models above, which we shall call two-dimensional particle dynamics, 
we consider simplified dynamical rules, which will be called one-dimensional:

{Dynamics $m=4$}: In this model only the $x$ coordinate of 
the particle matters and the particle is allowed to perform the 
jump $(x_p,y_p)\to(x_p',y_p')$
whenever the channel is open, {\em i.e.}~$h_{-}(x_p,t) h_{-}(x_p',t)>0$.
Then the effect of the interface dynamics on the possibility 
of the particular jump is fully determined by $h_{-}(x,t)$ 
and diffusion is most directly controlled by the bubbles. 
In this case, in addition to the jump of the particle 
in the horizontal direction, there is, when needed, 
a move in the vertical direction over one 
lattice unit to keep the particle between the 
interfaces such that $h_2(x_p,t) \le y_p \le h_1(x_p,t)$. 
The attempt rate of the jump in this case is denoted by $\nu $. 
The effect of forced moves on waiting times of the particle 
in the continuous-time dynamics is discussed in more detail 
in Sec.~\ref{sec:ParticleSim}.


\section{\label{sec:Numerics}Numerical methods}

\subsection{\label{Nfold}Interface dynamics}
For the dynamics of the interfaces, which is the most time-consuming 
part of the numerics, we used the so-called $N$-fold \cite{nfold} 
algorithm in our continuous-time Monte Carlo simulations. 
In the $N$-fold algorithm the possible transitions are divided into $N$ 
classes according to their probabilities. After finding those classes, 
one finds all lattice points $x$, which belong to a certain class $j$. 
The next step is to calculate the set of time-dependent variables 
$ Q_{i} =\sum_{k=1}^{i\le N}n_{j_{k}}P_{j_{k}}$, where $n_{j_{k}}$ is 
the number of those lattice points which belong to the class $j_{k}$ and $P_{j_{k}}$ 
is the probability associated with $j_{k}$. The class $j$ of
the event, which will occur, is next determined by finding $j$ such that
$Q_{j-1}\le R< Q_{j}$, where $R$ is a random number with uniform
distribution in the interval $( 0,Q_{N} \rbrack$. After finding the
class, one randomly chooses a location (on $h_{1}$ or $h_{2}$) from this
class. The waiting time for something to happen
in the system consisting of the two interfaces then 
is $\Delta t_I=- \ln(R_{1})/Q_N$, where $R_{1}$ is a 
random number with uniform distribution in the interval $\lbrack 0,1)$.

\subsection{\label{sec:ParticleSim}Particle dynamics}
To restrict the dimension of the parameter space, in our simulations 
we only consider cases where all possible attempts (see Fig.~\ref{fig:figure1}) 
of the particle to jump (within a given dynamics $m$) occur 
at the same rate as described in Sec.~\ref{sec:Particles}. 
For simplicity and when possible, we shall denote by $\mu$ the 
total jump rate of a free particle, which is then determined 
by the parameters 
$\alpha$, $\beta$,  $\gamma$ and  $\nu$ 
relevant in each case. 
In the presence of the 
interfaces this will become the total attempt rate such that 
with rate $\mu$ the particle will try to jump in a direction 
chosen randomly from the list of allowed processes. 
The waiting time $\Delta t_{P} $ after which the particle tries to jump is 
drawn form the exponential distribution $\Delta t_{P}=-\ln(R_{2})/\mu$, 
where $R_2$ is another uniformly distributed random number 
in the interval $(0,1]$.

In the combined model containing both the interfaces and the particle, 
the particle does not affect the dynamics of the interfaces, which 
will evolve as described in the first paragraph above. 
To include the diffusing particle, we need two 
waiting times: the waiting time $\Delta t_I$ of the interface 
system and the waiting time $\Delta t_{P}$ of the particle. 
We then keep track of two times: First, the latest instant of 
real time $t_I$, when there was a move in the interface system. 
Second, the latest instant of time $t_P$, when there was either 
and attempt of the particle to jump or a forced move of the particle. 
Then, if $t_I + \Delta t_I < t_P + \Delta t_P$, the next event in the 
system will be a move of the interface, otherwise the next event will 
be an attempt to move the particle.

After this, there still are in the continuous-time dynamics two 
obvious ways to handle the forced moves (in dynamics $m$=1,2,3): 
(A) The clock of the particle remains intact in a forced move
or (B) its waiting time $\Delta t_P$ for the particle is updated
after it. We shall consider both choices since they produce quite
different results and both can be physically justifiable
from some microscopy. With this choice the dynamics of the 
combined model becomes defined.

\subsection{\label{sec:Quantities}Sampling of main quantities}

The sampling of the observables was done with a constant 
time interval after reaching the steady state from an 
initial configuration consisting of two completely disordered 
interfaces at a fixed distance from each other. 
With the $N$-fold method reaching the steady state for 
interface configurations turns out not to be very difficult 
but especially for sampling of diffusion quite long runs were 
required.

Typically of the order of $10^3$ independent runs 
were performed for the interfaces in such a way that 
there were $10^2$ particles diffusing (independently 
of each other) between the interfaces, the linear 
size of the simulation cell being $L =100$ (for the 
finite-size scaling studies mentioned in the text larger 
systems were used). The reason for this procedure is that 
the dynamics of the interfaces even with the $N$-fold 
algorithm is computationally the most time-consuming part 
of the simulation, so we used each sequence of interface 
configurations to produce many particle trajectories, 
the total statistics thus being of the order of $10^5$.

To characterize the statistical properties of the interfaces 
$h_1$ and $h_2$, we use their roughness or width \cite{Barabasi95} 
defined as
\begin{eqnarray}
\label{eq:width}
  W(f) =\sqrt{\bigl\langle|h_i(x,t)-\bar{h}_i(t)\rangle|^{2}\rangle}
\end{eqnarray}
Here $i=1,2$ and $\bar{h}_i(t)$ is the spatially averaged 
height of the interface configuration at time $t$, and the 
angle brackets denote ensemble average, {\it i.e.}~average 
over independent simulations.
The kink density $\bar{k}(f)$ is the density of those 
locations $x$, for which $h_i(x-1,t) \neq h_i(x+1,t)$ for each of the 
interfaces $i=1,2$ separately. 
This definition follows from the fact that a BCSOS interface even in the 
flat state has an intrinsic roughness, because $|h_i(x,t) - h_i(x\pm 1,t)| = 1$. 
In the notations for these quantities we suppress the 
dependence on system size $L$.

The bubble size distribution per site was sampled at the 
same instant of times as the width of the interfaces. It gives 
the probability that a randomly chosen location $x$ belongs to a 
bubble of size $\ell$, which is the length of the bubble in $x$ 
direction. We shall be interested in bubble size distributions 
$P_j(\ell,f)$, normalized such that only bubbles 
with $j \le \ell < L$ are taken into account, 
$\sum_{\ell=j}^{L-1} P_j(\ell,f) = 1$, so that 
they are related by
\begin{equation}
 P_j(\ell,f)=P_0(\ell,f)\,\Bigl[1-\sum_{\ell = 0}^{j-1}P_0(\ell,f)\Bigr]^{-1}.
\end{equation}
In this $\ell=0$ means that $h_-(x,t)=0$ for site $x$.
In the configurations, where the interfaces did 
not touch each other at all, the bubble size was recorded as a 
count in bin $\ell=L$. 
For certain purposes we also sample the bubble size distributions 
normalized for $j \le \ell \le L$, which we denote by $P_j^\ast(\ell,f)$, 
so that $\sum_{\ell=j}^{L} P_j^\ast(\ell,f) = 1$. 
In addition, we compute the (physical) average bubble size $\bar{\ell}(f)$ 
and also $\bar{\ell}^\ast(f)$ defined as
\begin{equation}
 \bar{\ell}(f) = \sum_{\ell = 0}^{L-1}\ell P_0(\ell,f)
 \ \ \ \ \ \ \ \ \ \ \ \ \ \ 
 \bar{\ell}^\ast(f) = \sum_{\ell = 0}^{L-1}\ell P_0^\ast(\ell,f). 
\end{equation}
The mean waiting time for the change of the bubble size 
we shall denote by $\bar{\tau}(\ell,f)$.

The diffusion coefficients, denoted here by $D_{\rm{obs}}$, were determined in 
the long-time regime $\langle [\Delta x(t)]^{2} \rangle > \bar{\ell}^2 $,
as a slope of the mean square displacement via \cite{Kehr87}
\begin{eqnarray}\label{eq:diffusion}
\left\langle [\Delta x(t)]^{2}\right\rangle  \sim 2D_{\rm{obs}}t,
\end{eqnarray} 
where $\Delta x(t) = x_p(t_0+t)-x_p(t_0)$ is the particle displacement during 
a time interval of length $t$. 
In the simulations, to reach the hydrodynamic regime not only for 
interfaces but for diffusion as well, we run the dynamics long enough so that 
the square root of $\left\langle [\Delta x(t)]^{2}\right\rangle$ is much 
larger than the bubble size.
Particle diffusion in the $y$ direction, on the other hand, 
is in the long-time limit simply controlled by interface wandering and 
the effective size-dependent diffusion coefficient related to that.

\vfill\eject

\subsection{\label{sec:Parameters}Overview of the parameters and their meaning}

Our full model for diffusion restricted by fluctuating interfaces 
contais thus the following variable parameters: 

(i) The driving parameter $f$, defined in Eq.~(\ref{eq:parameters}) 
of Sec.~\ref{sec:Interface}, drives the interfaces towards each other. 
For interfaces to be coupled in the steady state we must have $f>0$. 
For increasing $f$, the channel for diffusion becomes narrower and 
the lengths of the locally open paths, i.e.~the bubble sizes, get 
smaller and the time scale of fluctuations gets longer. 
For $f \le 0$ the interfaces would be driven apart of each other 
and diffusion between them would become an unrestricted random walk.

(ii) The second parameter is the generic jump attempt frequency $\mu$ of the 
diffusing particle. The possible jump directions (horizontal, vertical or 
diagonal, see also Fig.~\ref{fig:figure1}) are controlled by the parameter 
$m=1,2,3,4$ described in Sec.~\ref{sec:Particles}. The simplest 
choice is the one-dimensional jumps in the model with $m=4$, 
which is a good starting point for analytical arguments for 
the scaling of diffusion with the model parameters.

(iii) The particle jumps are restricted by the dynamic environment provided by 
the interfaces. The technical details of this are given in Sec.~\ref{sec:Particles}, 
but there are no additional parameters involved. However, there 
remains two physically reasonable ways to realize the occasions, 
when the interfaces would possibly move the particle. This is 
described in Sec.~\ref{sec:ParticleSim}: In scheme (A) the particle 
is considered in such cases to move together with an interface with 
its diffusive 'clock' left intact and in (B) its 'clock' is updated 
after the interface move. The physical meaning of these choices 
is considered in Sec.~\ref{sec:Conclusions}.

\section{\label{sec:ResultsInt}Results for interface dynamics} 

In this Section we first complement the study of Ref.~\cite{Juntunen07} 
to characterize the time evolution of the coupled interfaces 
themselves (without a diffusing particle) in the stationary state. 
In the observed dynamics, a few time scales of interest can be monitored to 
gain insight also into the behavior of diffusion between the interfaces 
to be discussed in the next Section.
The first time scale is the average time $T_{\rm int}$ 
elapsed between consecutive changes in the interface configuration.
The second one, denoted by $T_{\rm lat}$, is the average time 
scale over which a single location in $x$ direction 
stays within a bubble $\ell>0$ in the $x$ direction. 
Also the third one describes the behavior of the interface 
system only: It is the average waiting time $T^{(j)}_{\rm bub}$ 
for a change in the bubble size, 
\begin{equation}
T^{(j)}_{\rm bub}(f) =\sum^{L-1}_{\ell=j} P_j(\ell,f) \bar{\tau}(\ell,f),
\end{equation}
where $P_j(\ell,f)$ is the normalized bubble size distribution 
and $\bar{\tau}(\ell,f)$ is the corresponding average waiting time 
for something to happen for a bubble of size $\ell$, 
see Sec.~\ref{sec:Quantities}. 
To obtain more detailed information, we computed $T^{(j)}_{\rm bub}$ for $j=0,1,2$ 
because the bubbles of size $\ell=0,1,2$ control diffusion for large $f$.

\begin{figure}[b]
	\begin{center}
		\includegraphics[width=0.45\textwidth]{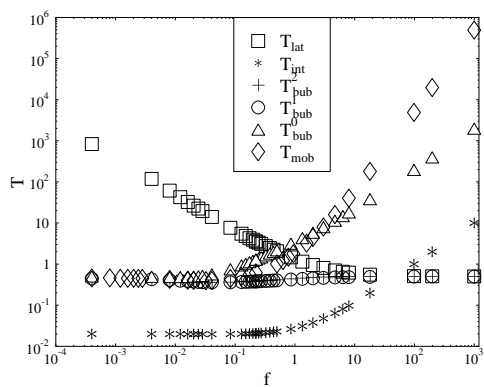}
	\end{center}
	\caption{Charasteristic timescales of the interfaces:
	the site-open time scale $T_{\rm lat}$, the interface-change 
	time scale $T_{\rm int}$, the bubble waiting times 
  $T_{\rm bub}^{(j)}$ for $j=0,1,2$ and 
  the effective mobility time scale $T_{\rm mob}$. 
		  }
\label{fig:figure2}
\end{figure}
\begin{figure}[b]
	\begin{center}
		\includegraphics[width=0.45\textwidth]{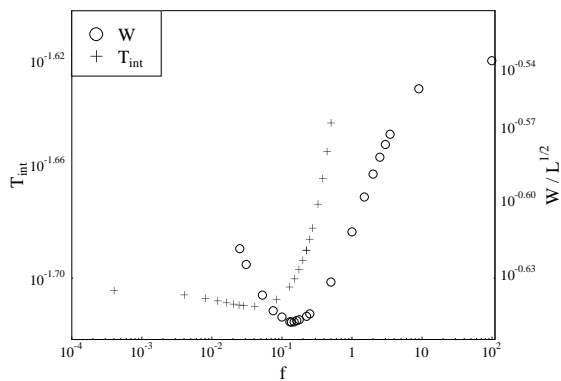}

	\end{center}
	\caption{The $T_{\rm int}$ and the scaled interfacial roughness $W$. 
	Note that the vertical scale is different for these quantities. 
	For this system size, the roughness dip is observed at $f = f_w \approx 0.15$. 
		  }
\label{fig:figure3}
\end{figure}

In Fig.~\ref{fig:figure2} we present the timescales $T_{\rm int}$, 
$T_{\rm lat}$, $T_{\rm bub}^{(j)}$ describing the interface dynamics. 
The time $T_{\rm int}$ is determined by the interface 
configuration and its behavior is not monotonic as can be seen 
from Fig.~\ref{fig:figure3}, where a shallow minimum is observed. 
On the other hand, the roughness $W$ as a function of $f$, 
shown in the same figure, displays a dip at $f_w \approx 0.15$. 
The non-monotonic behavior is explained by the entropic effects through 
the reduced configuration space available for the interfaces \cite{Juntunen07}. 
The finite-size scaling of the dip position, $f_w(L)$, 
was studied in Ref.~\cite{Juntunen07}, we only mention here the result 
$f_w \sim L^{-1/3}$ for large $L$. 
In passing we note that such a deroughening due to interactions between 
interfaces has been experimentally observed in another context, 
see the articles in Ref.~\cite{Bauer05}. 
It is evident from Fig.~\ref{fig:figure2} that the point $f=f_w$ is also the 
crossing point for the dynamical properties of the interfaces.
For $f>f_w$, $T_{\rm int}$ increases rapidly and displays the asymptotic 
behavior $T_{\rm int} \sim f$ for large $f$, because the rate-limiting factor 
is the time required to create new bubbles, which is proportional to 
$1/q_2 \sim (1-q_2)/q_2-1 \equiv f$ for $q_2 \ll 0$ in this limit.

In Fig.~\ref{fig:figure2} we also observe that $T_{\rm lat} \sim 1/f$ 
for $f \ll f_w$. This is controlled by the timescale of (finite) interfaces 
wandering apart from each other and then back together (effectively biased 
diffusion), which is proportional to $1/(p_2 - q_2) = 1/(2p_2-1) \sim 1/f$. 
For large $f$ we have $T_{\rm lat}\to 1/2$, because the smallest  
bubbles $\ell=1$ in this limit disappear with the rate $2q_1\to 2$. 
For increasing $f$, the time scale $T_{\rm bub}^{(0)}$ increases 
without limit with the waiting time for a bubble to appear
as $T_{\rm bub}^{(0)} \sim f$, but 
$T_{\rm bub}^{(1)} \to 1/2$ by the same argument as for $T_{\rm lat}$. 
Also the behavior of $T_{\rm bub}^{(2)}$ 
has the same asymptotics, as can easily be seen by inspecting 
the one and only possible shape of bubbles of size $\ell=2$, 
the rate-limiting factor being the shrinking of the bubble 
from its either end.

The last time scale shown in Fig.~\ref{fig:figure2} is the 
average waiting time for a change involving a bubble of 
size $\ell \ge 2$ to occur in a given lattice site,  
which we denote by $T_{\rm{mob}}$, since it is related to 
configuration changes that change the effective mobility 
of a particle by increasing or decreasing its possible range 
of motion. This differs from $T_{\rm bub}^{(2)}$ in that, 
for example, for a particle sitting at site $x$, where the 
sites belongs to a bubble with $\ell \le 1$, {\em i.e.}~for
a particle stuck in a locally closed configuration, in averaging 
$T_{\rm{mob}}$ we count the time for the particle to be mobile again, 
{\em i.e.}~for the site to be in a bubble with $\ell \ge 2$ again.
For a bubble with $\ell>2$, on the other hand, each change of its 
size will result in a greater or smaller effective mobility for 
the diffusing particle. For $f\gg f_w$ we have $T_{\rm{mob}} \sim f^2$, 
since the rate-limiting process contributing to it in this limit is 
a two-step process, where a bubble of size two becomes created 
starting from a configuration where there are no bubbles.

\begin{figure}[t]
	\begin{center}
		\includegraphics[width=0.45\textwidth]{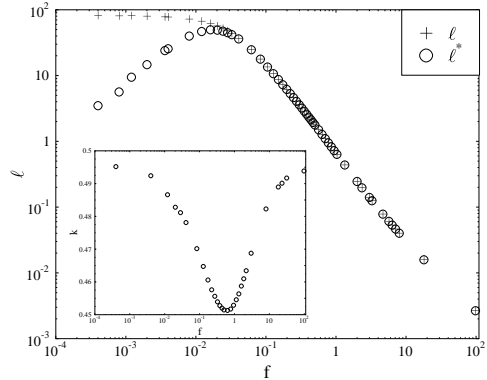}
	\end{center}
	\caption{The average bubble size $\bar{\ell}$ and kink density $\bar{k}$ as a function of $f$. 
	We show the bubble size normalized in two ways: the real $\bar{\ell}$ where bubble sizes 
	$0\le\ell < L$ and $\bar{\ell}^\ast$ where $0\le\ell\le L$ are taken into account. 
		  }
\label{fig:figure4}
\end{figure}

To characterize the properties of the interface configurations in more detail, 
we present in Fig.~\ref{fig:figure4} the mean bubble size $\bar{\ell}$ and the kink 
density $\bar{k}$. The non-monotonicity of the differently normalized (see the caption) 
mean bubble size $\bar{\ell}^\ast$ is a consequence of the fact 
that for finite $L$ the maximum size of a bubble is limited.
Below the roughness dip, for $f<f_w$, the probability of completely open bubbles 
($\ell \ge L$), neglected in the computation of $\bar{\ell}$, rapidly 
increases for decreasing $f$, while the number of such bubbles is 
essentially zero for $f > f_w$. 
In the vicinity of the dip, for $f \approx f_w$, we observe 
$\bar{\ell} \sim f^{-4/3}$, but for large $f$ it tends to 
$\bar{\ell} \sim f^{-1}$, as expected.
The kink density has a clear minimum slightly above $f_w$, 
which is consistent with the reduction of configuration space 
due to the interaction between the interfaces, resulting in more 
hill tops and valley bottoms, which are not kink sites. 
Note also that $\bar{k}(f\to 0) = \bar{k}(f\to \infty)$, 
because in the latter limit the interfaces are bunched 
together and their dynamics then reduces to that of a 
single isolated interface \cite{Juntunen07}.

\section{\label{sec:ResultsDif}Results for particle diffusion} 

For a continuous-time unbiased random walk the one-dimensional mean square 
displacement is of the form \cite{Kehr87}
\begin{eqnarray}\label{eq:variance}
\langle [\Delta x(t)]^{2} \rangle = \langle N_s\rangle(t) \sigma^{2}_{\Delta x},
\end{eqnarray}
where $\langle N_s\rangle (t)$ denotes the average number of jumps in time $t$, 
and $\sigma^2_{\Delta x}$ is the variance of the displacement of individual jumps. 
In what follows we shall use Eq.~(\ref{eq:variance}) to justify a theoretical 
model for diffusion between the interfaces for $f>f_w$. 
For the different models of particle dynamics in the presence of the interfaces, 
{\em i.e.}~the models $m=1,...,4$ defined in Sec.~\ref{sec:Particles}, we shall report our 
results as a function of the drive parameter $f$ and the total jump rate $\mu$ as 
the dimensionless ratios 
\begin{eqnarray}\label{eq:relative}
D_{m}(f,\mu)=\frac{D_{\rm{obs}}^{(m)}(f,\mu)}{D_{\rm{free}}^{(m)}(\mu)},
\end{eqnarray} 
where $D_{\rm{obs}}^{(m)}(f,\mu)$ is the observed diffusion coefficient according to Eq.~(\ref{eq:diffusion}) 
and $D_{\rm free}^{(m)}(\mu)$ is that corresponding to the same intrinsic 
jump rates without the interfaces. 
These ratios then give the effect of the interfaces on diffusion, while $\mu$ 
is the total attempt rate of jumps for the given model in units of the total 
attempt rate of local interface configuration changes.
For a free random walk with $ \sigma^{2}_{\Delta x} =1$, 
combining Eqs.~(\ref{eq:diffusion1}) and (\ref{eq:variance}) 
gives the diffusion coefficient for free motion in $x$ direction, for example, as 
$D_{\rm{free}}^{(1)}={\alpha}=\mu/4$ and $D_{\rm{free}}^{(4)}={\nu}=\mu/2$. 
We shall begin our discussion with the simplified dynamics $m=4$,
because for it the generic features of diffusion between the interfaces 
become more transparent.

\subsection{\label{sec:Diffusion1d}
Diffusion for the one-dimensional particle dynamics (model 4)}

\begin{figure}[b]
	\begin{center}
		\includegraphics[width=0.45\textwidth]{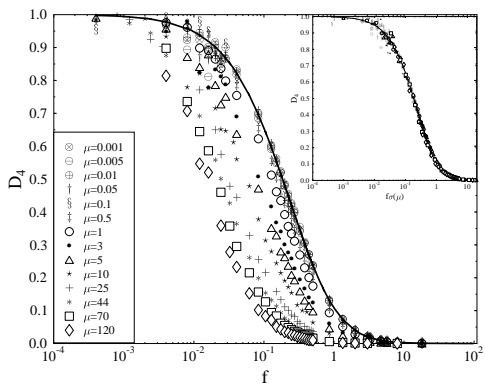}
	\end{center}
	\caption{Simulation results (plotting symbols) for $D_{4}(\mu,f)$ 
	 and the mean-field prediction for the small $\mu$ limit (solid line). 
	 In the inset we show the data collapse discussed in text.}
\label{fig:figure5}
\end{figure}

In the case of {\it slow particles} the interface configuration 
will change many times between the jump attempts of the particle.
In this case, the success ratio of the jump attempts can be evaluated 
from the bubble size distribution $P^\ast_0(\ell,f)$ and it is
\begin{equation}
 \label{eq:success}
 g(f) = \sum^{L-1}_{\ell=2} P^\ast_0(\ell,f)\frac{\ell-1}{\ell} +P^\ast_0(L,f),
\end{equation}
because only within bubbles with $\ell \ge 2$ the particles can move and 
with probability $(\ell-1)/\ell$ is an attempted jump possible since 
the particle cannot jump out of the bubble so that $2$ of the $2\ell$ 
attempts are blocked by the bubble edges.
The mean-field prediction for the diffusion coefficient is then
\begin{equation}
 \label{eq:mf}
 D_{\rm{mf}}^{(m)}(f,\mu)=g(f)\,D_{\rm{free}}^{(m)}(\mu).
\end{equation}
The simulation results for the model $m=4$ together with this mean-field approximation 
are presented in Fig.~\ref{fig:figure5}. 
For $\mu \ll 1/T_{\rm{int}}$, i.e.~for slow diffusion, the curves $D_4(f,\mu)$ 
follow the mean field prediction of Eq.~(\ref{eq:mf}).

\begin{figure}[b]
	\begin{center}
	\includegraphics[width=0.45\textwidth]{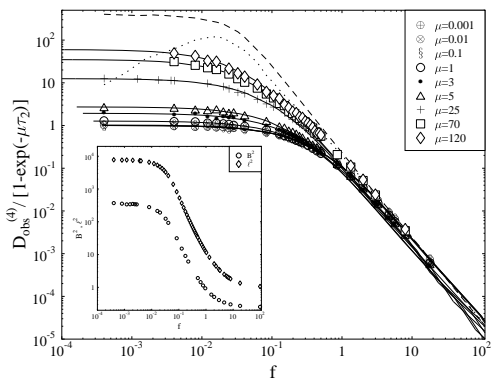}
	\end{center}
	\caption{Data collapse for the large $f$ behavior of $D^{(4)}_{\rm obs}$
	by using the scaling form of Eq.~(\ref{eq:tau}).
	The plotting symbols show the simulation results and 
	the solid lines are the mean-field approximation for small $\mu$ 
	of Eq.~(\ref{eq:mf}) with $f$ scaled by the factor $\sigma(\mu)$ 
	given in Eq.~(\ref{eq:sigma}). 
	The dashed line is the adiabatic approximation of Eq.~(\ref{eq:adiab})
	and the dotted line is the finite-size-corrected infinite-rate 
	approximation described in text.
 	In the inset we show the comparison of the behavior of the jump-length 
	factor $B^2$ and the mean squared bubble size $\ell^2$.}
\label{fig:figure6}
\end{figure}

In the inset of Fig.~\ref{fig:figure5} we show a data 
collapse by using for $f$ the scaling factor
\begin{equation}
 \label{eq:sigma}
 \sigma(\mu) = \sqrt{1+c\mu} 
\end{equation}
with $c=1$. 
This is an interpolation of the large $\mu$ behavior 
$\sigma(\mu)\sim \mu^{1/2}$ for $\mu \gg 1$, with the 
interface and particle time scales well separated, 
and the small $\mu$ behavior $\sigma(\mu) \approx 1$ 
for $\mu \ll 1$, which is exactly the mean-field result 
discussed above. With this scaling, our simulation data for $D_4(f,\mu)$ 
for $f < f_w$ is nicely collapsed onto the slow-particle curve 
defined by Eq.~(\ref{eq:mf}) and even better with the 
choice $c \approx 1.1$ (not shown), which we assign to 
finite-size effects. 
However, to obtain a good data collapse for $f > f_w$ we need 
a different scaling combination, 
\begin{equation}
 \label{eq:tau}
 D_{\rm obs}^{(m)} \ \to \ D_{\rm obs}^{(m)}/[1-\exp(-\mu\tau_2)], 
\end{equation}
corresponding to an exponential clock 
with $\tau_2=T_{\rm{bub}}^{(2)}(f\!\to\!\infty) = 1/2$. 
This scaling form results from the characteristic time scale 
of smallest bubbles allowing diffusion of particles in this regime. 
This is controlled by the driving parameter $f$ such that for 
$f \gg 1$ there are mainly bubbles bubbles of size $\ell = 1$, 
but for diffusive jumps to take place bubbles of size $\ell = 2$ 
are needed (with the time units chosen, once created, they stay 
open at least over one unit of time, cf.~Sec.~\ref{sec:ResultsInt}). 
This scaling is demonstrated 
in the loglog plot of Fig.~\ref{fig:figure6}. Because of the 
different rate-limiting mechanisms these scaling forms, 
below and above the dip region, are incompatible.

Next we consider the case of {\it fast particles}. 
If the particle jump rate is fast compared to the interface
dynamics, an adiabatic approximation can be done as follows. 
The particle is then trapped inside a bubble and its 
location becomes uniformly distributed in the timescale 
of the bubble dynamics and the (effective) location of 
the particle can change only when the size of the bubble changes. 
The length of an effective particle jump via this mechanism 
is approximately the length of the displacement of the 
center-of-mass location $b$ of the bubble. By applying 
Eqs.~(\ref{eq:diffusion},\ref{eq:variance}) we then obtain
\begin{equation}
 \label{eq:adiab}
 D_{\rm adiab}(\mu,f)= 
 \frac{1}{2}\,\frac{1}{T_{\rm mob}}\,B^2(f),
\end{equation}
where $T_{\rm mob}$ is the time scale of bubble motion 
and $B^2(f)$ is the mean-square displacement 
(per jump) of the bubble obtained from 
\begin{eqnarray}\label{eq:diffusion1}
 B^q(f) = \langle | b_{\rm new} - b_{\rm old} |^q \rangle,
\end{eqnarray}
with $q=2$. 
Here $b_{\rm new}$ and $b_{\rm old}$ are the locations 
of a bubble before and after the change in the bubble size.
This approximation is shown by the dashed line 
in Fig.~\ref{fig:figure6}. 
For large $f$ the expected behavior $D \sim f^{-2}$ is recovered, 
since the bubble jump length and thus $B^2$ becomes a constant 
for isolated bubbles and $ T_{\rm mob} \sim f^2 $.
For $f>f_w$, the bubble motion is dominated by increasing 
or decreasing the bubble size by one lattice unit and, 
as expected, for fast particles the adiabatic 
approximation works well.
For $f<f_w$, merging and dissociation of bubbles becomes 
important. 
We also tested the choice $q=1$, with the mean jump length 
squared, giving less weight to long jumps and thus giving a smaller 
diffusion coefficient for 
small $f$ (for large $f$ the smallest jumps dominate in any case), 
but the approximation is not essentially better.

The dotted curve just below the adiabatic approximation shows an approximation 
obtained by assuming an infinite jump rate of the particle such that after each 
interface configuration change a new location for the particle is 
drawn evenly distributed inside a bubble, but such that the particle 
displacement in the case of a completely open interface configuration 
is restricted by the system size. A possible way to extend this 
approximation for smaller $f$ would be to utilize the known form 
of the bubble-size distribution for large $\ell$. 
This way we obtain a reduction of the diffusion coefficient 
that seems to work around $f=f_w$ but apparently fails 
for small $f$, {\em c.f.}~the behavior of $\ell^\ast$ 
in Fig.~\ref{fig:figure4}.
Another natural approximation would be to consider the 
size of the bubble as defining an effective mobility of 
the diffusing particle. Qualitatively, the behavior 
of $B^2$ and the mean squared bubble size 
$\ell^2 $ are quite similar, but with 
incompatible limits for small and large $f$.
Unlike in Eq.~(\ref{eq:adiab}) for $B^2$, 
it turns out to be difficult to assign a natural 
rate factor to $\ell^2$ to develop a 
reasonable approximation for the diffusion coefficient.

\begin{figure}[h!]
	\begin{center}
		\includegraphics[width=0.45\textwidth]{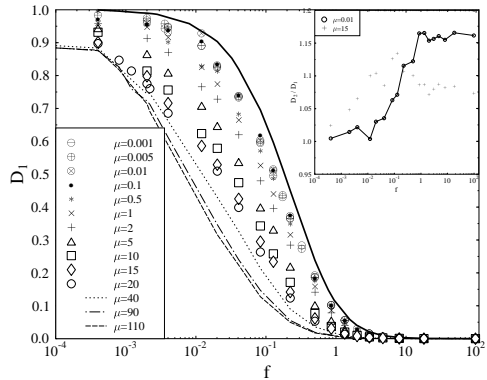}
	\end{center}
	\caption{Diffusion coefficient $D_1$ for the particle clock updating scheme A. 
	For comparison we show also 
	by the full line the mean-field approximation, c.f.~Fig.~\ref{fig:figure5}.
	In the inset we show the ratio $D_2(\mu/2)/D_1(\mu)$.}
\label{fig:figure7}
\end{figure}

\begin{figure}[htbp!]
	\begin{center}
		\includegraphics[width=0.45\textwidth]{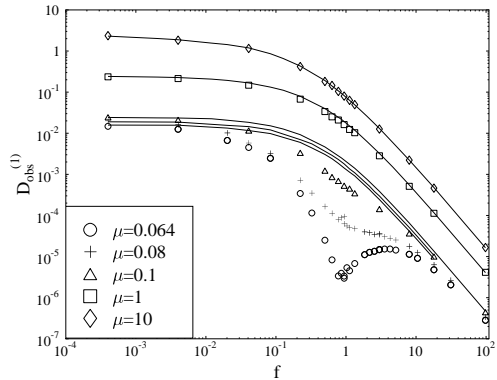}
	\end{center} 
	\caption{Diffusion coefficient $D_1$ for the particle clock updating scheme B. 
	The full curves show the corresponding data for scheme A.}
\label{fig:figure8}
\end{figure}

\subsection{\label{sec:Diffusion2d}Diffusion for the two-dimensional particle dynamics (models 1,2,3)}

We next consider the effect of the 'microscopic' dynamics on diffusion. 
First, in Fig.~\ref{fig:figure7} we show the diffusion coefficient 
for dynamics $m=1$ with the pushes of the particle by the interfaces 
handled according to clock-updating scheme (A) and in Fig.~\ref{fig:figure8} 
according to scheme (B), see Sec.~\ref{sec:Particles} for details. 
Now it is not just the length of the bubble but also the shape of it 
that affects diffusion. In (A) the 
particle waiting time is updated after each jump attempt, in (B) it 
is updated also after each push by the interfaces. The difference 
between (A) and (B) appears for long particle waiting times (for small $\mu$) ´
in the regime where there are frequent pushes, {\em i.e.}~the drive $f$ 
needs to be strong enough, but not so strong that diffusion would be 
completely dominated by the motion of small bubbles as discussed above. 
The coincidence of the waiting times of the interface motion 
and the particle jump attempts produces at a finite value of $f$ 
in (B) behavior that looks 
like the suppression of diffusion sometimes observed close to phase 
transitions. The dip in $D_1$ appears close to the dip in the kink 
density $k$ (see the inset of Fig.~\ref{fig:figure4}), where in the 
local interface configurations there are more sites with possible 
interface configuration changes (at a kink site the interface 
is locally frozen) and thus more pushes. 
Other variants of the jump rate modification are 
conceivable, {\em e.g.}~physically it might happen that the contact 
with the interfaces could boost the particle jump rates, but in 
the region of the parameter space, where bubble dynamics is the 
rate-limiting factor, it would not essentially change diffusion 
from what what is observed in scheme (A).

The corresponding mean-field approximation of Eq.~(\ref{eq:mf}) shown 
in Fig.~\ref{fig:figure7} by the full curve is presented to help the 
comparison between $m=1$ and $m=4$.
As evident from the comparison of Fig.~\ref{fig:figure7} 
and Fig.~\ref{fig:figure5}, scaling by the factor $\sigma(\mu)$ 
of Eq.~\ref{eq:sigma} would not yield a very good data collapse 
for $f \ll f_w$ in the case of two-dimensional particle dynamics. 
For particles fast both in the horizontal and vertical direction, 
the blocking of diffusion by 'collisions' with the interfaces is quite 
efficient and leads to a considerable reduction of $D_1$ in this regime. 
Note that even for small $f$ we have $D_1<1$ for fast particles, 
since particle diffusion in the vertical direction is faster than 
that of the interfaces for any $f>0$. 
This in part explains the spreading of the curves $D_1(f)$ for $f<f_w$.
In the inset of Fig.~\ref{fig:figure7} we show also the difference 
between the particle dynamics $m=1$ and $m=2$ with the clock updating scheme A 
(concerning dynamics $m=3$, we find that $D_1 \le D_3 \le D_2$). 
For reasonable comparison, the jump rates for each dynamics have 
been chosen such that the total jump rates in the $x$ direction match, 
thus we plot $D_2(\mu/2)/D_1(\mu)$. 
Due to the limited accuracy of $D_m$, the ratio $D_2/D_1$ becomes 
somewhat noisy, but a few general observations can be made. 
A difference first develops for $f>f_w$, because diffusion becomes 
more restricted by the interfaces within bubbles of the type seen 
on the right ($x=17,...,20$) in the snapshot of Fig.~\ref{fig:figure1} becoming 
prevalent. In such interface configurations, many of the 'horizontal' 
jumps (for $m=1$) become blocked and diffusion along the narrow channel 
requires also 'vertical' jumps, while the diagonal jumps (for $m=2$) 
are more effective for particle transport. 
However, for $f\gg f_w$ diffusion again becomes dominated by bubble motion 
the way it was for $m=4$, and $D_m \sim f^{-2}$ for $m=1,2,3$ for both particle 
clock updating schemes (A) and (B) as seen in Fig.~\ref{fig:figure8}.

\section{\label{sec:Conclusions}Discussion} 

To summarize, we have considered one and two dimensional 
continuous time random walkers constrained between two 
evolving interfaces symmetrically driven towards each other.
A surprisingly complicated phenomenology appears. 
First, in the interface model itself there is a dip 
in the interface roughness at a finite value $f_w$ 
of the parameter $f$ describing the drive of the 
interfaces against each other \cite{Juntunen07}, 
with different kinds of interface and bubble dynamics 
below and above the dip. 
The crossover from bubble-dominated (interface-dominated) 
to almost free diffusion is controlled by the relative 
jump rate of the particle and its interplay with the 
rate of the interface time evolution.
In the analysis of this crossover, simple scaling arguments 
for the two regimes are incompatible, both 
physically and formally, so that the full 
behavior of diffusion cannot be described by a single scaling form. 
This can be expected to be a generic property of transport 
restricted by fluctuating interfaces.

Furthermore, diffusion was found to depend on the microscopic 
details of the interaction between the interfaces and the 
diffusing particles. In particular, its immediate effect 
on the waiting time of particle jumps was shown to be considerable, 
especially for dynamics physically corresponding to diffusion 
on a lattice in the large-friction limit, which can be 
realized at domain boundaries in adsorption systems. 
In a spatial continuum this effect would be absent, 
as would be the roughness dip. However, 
the finite-size case, with the underlying microscopic structure 
not washed out in coarse graining, can be of interest in its own 
right in nanoscale applications. Then also the size of the channel for 
the particles will induce a relevant length scale, in addition to 
the length scale related to the diffusive jumps (lattice) and 
the one related to the environment (bubbles).

In this work we have considered diffusion that is effectively 
one-dimensional even if the diffusive jumps and interactions 
between the tracer particles and the interfaces result from 
two-dimensional dynamics. The environment for the diffusing 
particles is then described by a chain of bubbles, where as in 
higher dimensions more complicated topologies (networks) would 
arise, in some cases leading to a percolation problem. 
In such studies, like in the present one, a considerable 
problem is the wide gap between the 'microscopic' timescales 
related to the dynamics of the particles and the environment, 
and the 'hydrodynamic' timescale corresponding to diffusion 
over length scales larger than any structures in the environment 
experienced by the particles. We hope our work would inspire 
further theoretical and experimental studies of diffusion in 
evolving and constraining environments.

\begin{acknowledgments}
We thank Dr.~Otto Pulkkinen for useful discussions.
\end{acknowledgments}

\vfill\eject


\end{document}